\documentclass[10pt, aps, physrev, twocolumn, superscriptaddress]{revtex4-2}
\usepackage[colorlinks=true, linkcolor=blue, citecolor=blue, urlcolor=blue]{hyperref}
\usepackage{amsmath,amsfonts}
\usepackage{braket}
\usepackage{graphicx}
\usepackage{xcolor}
\usepackage{comment}

\begin{document}

\title{Effective Three-Boson Interactions using a Separable Potential}

\author{Corinne Beckers}
\email[]{corinne.beckers@uantwerpen.be}
\affiliation{Theory of Quantum Systems and Complex Systems, Universiteit Antwerpen, B-2000 Antwerpen, Belgium}

\author{Jacques Tempere}
\affiliation{Theory of Quantum Systems and Complex Systems, Universiteit Antwerpen, B-2000 Antwerpen, Belgium}

\author{Jeff Maki}
\affiliation{Department of Physics, University of Konstanz, 78464, Konstanz, Germany}

\author{Denise Ahmed-Braun}
\affiliation{Theory of Quantum Systems and Complex Systems, Universiteit Antwerpen, B-2000 Antwerpen, Belgium}

\date{\today}

\begin{abstract}
Effective field theories (EFTs) are widely used to study many-body systems by describing two-body interactions using zero-ranged contact potentials. However, when extended to three-body processes, these contact interactions lead to divergences due to the absence of an intrinsic length scale. 
In EFT, this is typically resolved by introducing a zero-ranged three-body interaction, which can be renormalized to make the low-energy physics independent of the short-distance physics.
However, when the two-body potential has a finite range, such as in separable potentials, there is no need for such renormalization. In this work, we derive the integral equation for the three-body scattering amplitude for separable potentials, and solve it in the strongly-interacting regime. 
With our model, we retrieve the known analytic form of the scattering amplitude for inelastic scattering processes and formulate a new scaling law for elastic three-body scattering processes.
\end{abstract}

\maketitle

\section{Introduction \label{section: introduction}}

In dilute ultracold atomic gases, interactions are typically modeled as two-body processes since the microscopic interactions between atoms are pairwise. In the strongly-interacting regime however, it is known that these two-body interactions can result in non-trivial \textit{effective} many-body interactions~\cite{Yamaguchi1979, Platter2009,  Wang2012, Hammer2013, Endo2024}, the simplest  of which are three-body interactions. 

Ultracold quantum gases are particularly well-suited to study effective three-body interactions. In quantum gases, generally the interactions between two particles are controlled by only s-wave scattering~\cite{ Pitaevski2003, Pethick2008}, described by the s-wave scattering length $a$ (with small corrections due to the effective range $r_0$). These interactions can be experimentally controlled using Feshbach resonances~\cite{Moerdijk1995, Pitaevski2003, Ketterle2008, Bloch2008, Chin2010}, allowing for effective many-body interactions to occur at higher probability.

Of particular interest is the unitary limit of infinitely strong interactions, where $1/a\rightarrow 0$. It is well known that for such strongly interacting Bose gases at unitary the Efimov effect becomes relevant~\cite{Efimov1970energy,efimov1970weakly, Hammer2010, Naidon2017, Greene2017}. This effect describes how an infinite series of three-body trimer states with geometric scaling form when the dimer-state is at zero-energy \cite{Efimov1970energy, efimov1970weakly}. The Efimov effect is experimentally detectable in atomic Bose gases, and has been shown to be important for describing enhanced three-body recombination \cite{Burt1997, Kraemer2006, pollack2009, Zaccanti2009, Berninger2011,Dyke2013, Shotan2014,Dincao2015}, the growth and presence of three-body correlations \cite{Braaten2003, Haller2011, Barth2015, Fletcher2017, Eigen2017, D’Incao2018, Colussi2018,Colussi2020}, obeying discrete scaling and universality \cite{pollack2009, Zaccanti2009, Wang2012, Srensen2013, Pires2014, Huang2014, Langmack2018,Mestrom2019Finite}, and atom-dimer resonances \cite{Knoop2009, Ferlaino2011}.

For identical bosons, the scale invariant wavenumber for the $n^{\text{th}}$ trimer at unitarity, $\kappa^{(n)}_*$, differs from the next more deeply bound trimer state, $\kappa^{(n+1)}_*$, by the discrete scaling relation:
\begin{align}
        \kappa_*^{(n+1)} = e^{-\pi/s_0} \kappa_*^{(n)} \approx  \kappa_*^{(n)} /22.7 ,
        \label{eq:kappa_ratio}
\end{align}
where $s_0 = 1.00624$~\cite{Efimov1970energy, Braaten2006, Esry1999, bedaque1999, Braaten2001}. For contact (zero-range) potentials, the Efimov spectrum is unbounded from below and leads to the well-known Thomas collapse~\cite{Thomas1935, Naidon2017}. This collapse is absent for short-range potentials, where a three-body parameter that scales with the potential range naturally emerges and restricts the deepest bound trimer state to a finite trimer wavenumber $\kappa_0^*$. 

In many-body physics, finite-range potentials are often replaced with their zero-ranged counterparts, as the former are too difficult to handle in practice. Often it is more practical to use effective field theories (EFTs) with zero-range interactions, and cure divergencies by matching the low-energy physics to physically observable quantities~\cite{WEINBERG1979, Georgi1993, Bedaque1999renormalization, Burgess2007, Platter2009, Hammer2020}. This procedure has been very successful in the analyses of few-body processes in many-body backgrounds, where contact interactions can serve as useful approximations to more complex (van der Waals-type) potentials~\cite{Lieb1963, Pitaevski2003, Pethick2008, Bloch2008, Chin2010}.
However, it is important to note that the correct renormalization of two-body processes in EFTs \textit{does not} automatically result in the correct renormalization of three-body (and higher order) processes. Instead, the EFT with renormalized two-body interaction strength $g_2$ will have to be explicitly altered to include an effective three-body contact interaction term with strength $g_3$ that renormalizes effective three-body processes.

In this work, we present an alternative approach to the EFT treatment by considering a many-body theory where the underlying two-body interactions are modeled using a simple finite-range separable potential with (step-function) form factors. The explicit introduction of a finite range cures the presence of divergences at the level of both two- and three-body processes (and higher order processes) automatically, without the need of renormalization and the introduction of an effective three-body interaction parameter $g_3$. 
We compute the three-body scattering amplitude in the unitary limit using both the EFT framework as outlined in Ref.~\cite{Braaten2006}, as well as by using our theory with separable interactions. We show how we can correctly retrieve the known analytical form of an effective $g_3$ for inelastic scattering at unitarity in the contact limit, and extend these results to describe elastic three-body scattering. We furthermore show how the effective range is manifest in the phase of the log-periodic oscillations in the scattering amplitude.
 
Our paper is organized as follows. In section~\ref{section: general framework}, we summarize two-body scattering for separable potentials.  Next we apply this approach to three-body scattering, and introduce the Alt-Grassberger-Sandhas (AGS) formalism in Sec.~\ref{section: three-body equations}. In Sec.~\ref{sec:three-body}, we explicitly calculate the three-body scattering amplitude using the separable potential and compare it to EFT. Section~\ref{section: comparative study} examines scattering amplitudes to determine the Efimov spectrum, and both the inelastic and elastic scattering amplitudes. Finally, in Sec.~\ref{section: conclusion}, we present our conclusions and outlook.

\section{Two-Body Scattering with Separable Potentials \label{section: general framework}}

First we consider the case of two identical particles interacting via a separable potential. For simplicity we assume the pairs of particles are in the center of mass frame. The dynamics of the relative motion is governed by the Hamiltonian
\begin{equation}
    H = \frac{{\bf k}^2}{2\mu}  + V({\bf r}),
\end{equation}
where $\mu$ is the reduced mass, $\bf k$ is the relative momentum, and $V(\mathbf{r})$ is a central potential that depends on the relative coordinate $\mathbf{r}$ between the two-particles.
The exact form of the potential $V(\mathbf{r}) = \langle {\bf r} | V | {\bf r} \rangle $ is generally complicated for neutral atoms; it is typically characterized by a Lennard-Jones type profile, featuring a steep short-range Pauli repulsion and a long-range van der Waals attraction~\cite{Jones1924, Stone1996}. The eigenstates of the free part of this Hamiltonian can be represented as plane waves where we choose the normalization such that: \mbox{$\braket{{\bf x} | {\bf k}} =e^{i {\bf k \cdot x}}/(2\pi)^{3/2}$}.

However, the low-energy scattering properties of the system are mostly insensitive to the short-range details of the potential. As such, we limit ourselves to the use of a separable potentials. Generally, separable potentials can be constructed by following the approach developed by Ernst, Shakin and Thaler (EST), for a two-particle system in three dimensions~\cite{Ernst1973}.
This method begins with the EST form of the two-body potential
\begin{align}\label{vsep}
    V_{sep} = \sum_{ij}V \ket{\psi_{{\bf k}_i}} g_{ij}\bra{\psi_{{\bf k}_j}}V,
\end{align}
where the sum runs over the different eigenenergies $E_i$ and $E_j$, $g_{ij}$ is a constant.
This potential is defined in such a way that the Hamiltonian with the separable form reproduces the same eigenenergy for an eigenstate $\ket{\psi_{\bf{k}}}$ with momentum ${\bf k}$, as the Hamiltonian with the original interaction $V({\bf r})$. 
In this work, we restrict ourselves to the simplest version of the EST potential by retaining only a single term in the above expansion. This simplification preserves the essential feature of introducing a finite-range to our potential.
In this case, the separable potential reduces to 
\begin{align}\label{seppot}
    V_{sep}&= g \,\ket{\xi}\bra{\xi}.
\end{align}
Here we define the form factors as $\ket{\xi}=V \ket{\psi_{k_i}}$ and the interaction strength as $g = 1/\braket{\psi_\textbf{k}| V|\psi_\textbf{k}}$. 
The expression for the form factors in momentum space can be obtained from the partial wave expansion for a short-range potential~\cite{Sakurai2020, Ahmed-Braun2023}. 
For s-wave interactions, the form factors become constant in the low-energy limit~\cite{Mestrom2021}. Consequently, the separable approximation on the potential yields
\begin{align}
        \braket{{\bf k}|\xi} \underset{k \rightarrow 0}{=} C(\Lambda)+ \mathcal{O}(k^2),
\end{align}
where $C$ is a constant that depends on the cutoff $\Lambda$. This potential can then be matched to the scattering parameters so that the cutoff corresponds to the physical range of the interaction. For simplicity, we assume a form factor that is a  step function
\begin{align} \label{eq:form_factor_kspace}
    \braket{\textbf{k}|\xi} = \xi(k)= \Theta(\Lambda-|{\bf k}|), 
\end{align}
where $ \Theta(k)$ represents the Heaviside step function. \par

All the two-body scattering properties can be encapsulated in the two-body $T$-matrix, defined through $T\ket{\bf k} = V\ket{\psi,+}$, where $\ket{\psi,+}$ is the outgoing two‑body scattering wave function in the presence of the interaction potential $V$ \cite{Sakurai2020}. The $T$-matrix elements $T({\bf k'},{\bf k})=\langle{\bf k'} | T | {\bf k}\rangle$ describe the probabilities for a pair of particles with no center-of-mass momentum and a relative incoming momentum $\mathbf{k}$ to scatter into an outgoing state with the same center-of-mass momentum and a relative momentum $\mathbf{k}'$. The computation of this matrix can be carried out using the Lippmann-Schwinger (LS) equation~\cite{Sakurai2020}
\begin{equation}\label{LS}
    T = V + VG_0T,
\end{equation}
where $G_0 = (E-H_0+ i\hbar \varepsilon)^{-1}$ denotes the free Green's function for two particles in the center-of-mass frame.

If the bare interaction is separable and of the form in Eq.~\eqref{seppot}, the full solution for the two-body $T$-matrix will also be separable
\begin{align}
\braket{\mathbf{k}'|T|\mathbf{k}} = \tau(E)\,\Theta(|\Lambda-\mathbf{k}|)\Theta(|\Lambda-\mathbf{k'}|),
\end{align}
where we have used Eq.~\eqref{eq:form_factor_kspace} to specify the form factor. The function $\tau(E)$ only depends on the relative energy of the two-particles, while the dependence on the ultraviolet cutoff, $\Lambda$, is left implicit. Substituting the separable form of the $T$-matrix into Eq.~\eqref{LS} gives the following equation for $\tau(E)$
\begin{align}
\tau^{-1}(E) = \frac{1}{g} - \int d\mathbf{k}
'' \frac{\Theta(|\Lambda-\mathbf{k}'' |)^2}{E-\mathbf{k}'' /2\mu + i \epsilon},
\end{align}
which can be solved analytically
\begin{align} \label{eq:tau_finite_Lambda}
    \tau^{-1}(E)& = \frac{1}{g} + 8 \pi\mu \Lambda \nonumber \\
    &- 8 \pi \mu \sqrt{-2\mu E - i \epsilon} \tan^{-1}\left(\frac{\Lambda}{-\sqrt{-2\mu E - i \epsilon}}\right).
\end{align}
We renormalize the theory by considering the two-particle scattering at low energies. To this end, we note that the $T$-matrix is related to the s-wave  scattering amplitude according to \cite{Sakurai2020}
\begin{equation}\label{fk}
    f({\bf k'},{\bf k}) = -\frac{4\pi^2\mu}{\hbar^2}T({\bf k'},{\bf k}).
\end{equation}
At low energies, the s-wave scattering amplitude can be expanded in terms of the relative energy using the effective range expansion \cite{Taylor1972}. By comparing the effective range expansion to the $T$-matrix in Eq.~\eqref{eq:tau_finite_Lambda} using Eq.~\eqref{fk}, one can show that the regularized form of the $T$-matrix is
\begin{align} 
    \tau^{-1}(E)&=4\pi^2\mu\left[\frac{1}{a}-\sqrt{-2\mu E-i\varepsilon}\right]+ \mathcal{O}(E/\Lambda), \label{T2body}
\end{align}
where
\begin{align}
    a &= \frac{4\pi^2 g \mu}{1+8\pi g \Lambda \mu}, \label{a}
\end{align}
which denotes the s-wave scattering length. This result is equivalent to~\cite{Braaten2006}, with a difference of normalization where $g \rightarrow g_2/4 \times(2\pi)^{-3}$.
Thus by setting the properties of the low-energy scattering $a$ and the range of the model $\Lambda$, the two-body scattering properties are completely regularized.

\section{Three-Body Scattering with Separable Potentials \label{section: three-body equations}}
\label{sec:three-body}
We next consider three-body scattering due to two-body interactions from a separable potential using the Alt–Grassberger–Sandhas (AGS) formalism~\cite{Taylor1972}. This formalism yields a set of three‑body transition operators that are linearly related to the three‑body scattering amplitudes. Our starting point is the three‑body Hamiltonian
\begin{align}\label{eq:H3}
    H^{(3b)} = H_0+ V_{ab}({\bf r}_{ab})+V_{ac}({\bf r}_{ac})+V_{bc}\left({\bf r}_{bc}\right),
\end{align}
for particles $a,b$ and $c$. Here $H_0$ denotes the sum of the kinetic energies of the three particles, and ${\bf r}_{ij} = {\bf r}_i-{\bf r}_j$. Since the interaction is only pairwise, we have three interaction terms. For simplicity we will label these terms from now on using a Greek index which identifies the spectator particle or the interaction channel~\cite{Taylor1972}. 

In this notation, the index $\alpha$ can take the values $\alpha = 0,1,2,3$, where  $\alpha=0$ denotes the channel in which all particles scatter and asymptotically approach free particles, while $\alpha = 1,2,3$ label the channels in which a dimer state is formed between a pair of particles. In Tab.~\ref{tab:1} we show how the the different channels are connected to the formation of dimers between the atoms $a$, $b$, and $c$.

\begin{table}[]
\begin{tabular}{|c|c|c|}
\hline
Channel ($\alpha$) & Dimer & Free atoms \\ \hline
0        &  n.a.  & $a$,$b$,$c$     \\ \hline
1        & $b$,$c$   & $a$         \\ \hline
2        & $a$,$c$   & $b$         \\ \hline
3        & $a$,$b$   & $c$         \\ \hline
\end{tabular}
\caption{channels in terms of which atoms form free operators, and which atoms form dimers. In channel $\alpha = 0$, all atoms are free, while for $\alpha \neq 0$ there is a single dimer between two of the three atoms.}
\label{tab:1}
\end{table}

The corresponding channel Hamiltonians are obtained by only keeping the interaction between the particles that are bounded together, $H^\alpha = H^{(3b)} - \sum^{'}V_{ij}$, where the summation runs over the two pairs of particles that are not bounded with each other~\cite{Taylor1972}. Finally, the scattering potential for a channel $\alpha$ is given by
\begin{align}
    V^\alpha = H^{(3b)} - H^\alpha.
\end{align}
The corresponding three‑body Lippmann–Schwinger equation for channel $\alpha$ takes the form
\begin{align}
\label{eq:LS3}
\ket{\psi,\alpha +} = \ket{\phi,\alpha}+GV^{\alpha}\ket{\phi,\alpha + }, 
\end{align}
where $\ket{\psi,\alpha +}$ is the outgoing scattering eigenstate of the full three-body Hamiltonian $H^{(3b)}$ as presented in Eq.~\eqref{eq:H3}, and $\ket{\phi,\alpha}$ denotes the eigenstate of the channel Hamiltonian. The three-body Green's function for channel $\alpha$, $\mathcal{G}$, satisfies its own LS equation 
\begin{align}
\mathcal{G}_{\alpha} = G_{\alpha} + G_{\alpha} V^{\alpha} \mathcal{G}_{\alpha}, 
\end{align}
where we have introduced the non-interacting, channel-specific Green's functions $G_{\alpha} = (E-H^{\alpha} + i\epsilon)^{-1}$.

\subsection{AGS Formalism  \label{subsection: AGS formalism}}
The three‑body Lippmann–Schwinger equation in Eq.~\eqref{eq:LS3} does not provide a unique solution to the three‑body scattering problem. In addition to the direct channel, one must also account for the possibility of rearrangement processes~\cite{faddeev2013}. A unique physical solution is obtained by exploiting the fact that all scattering states are eigenstates of the three-body Schrödinger equation. This naturally leads to a set of three coupled Faddeev equations~\cite{Faddeev1960, faddeev2013} whose sum reconstructs the total scattering wave function \footnote{The unique solution is given by 
\begin{align}
\ket{\psi,\alpha +} =  G_0 \sum_{\mu}^3 V_{\mu} \ket{\psi,\alpha +} = \sum_{n}^3 \ket{\phi,(\alpha \mu)}, 
\end{align}
with the Faddeev equations 
\begin{align}
\ket{\psi,(\alpha \alpha)} &= \ket{\phi,\alpha}+G_{\alpha}V_{\alpha}\left[\ket{\psi,(\alpha \beta)}+\ket{\psi,(\alpha \gamma)}\right] \\
\ket{\psi,(\alpha \beta)} &= G_{\beta}V_{\beta}\left[\ket{\psi,(\alpha \alpha)}+\ket{\psi,(\alpha \gamma)}\right] \\
\ket{\psi,(\alpha \gamma)} &= G_{\gamma}V_{\gamma}\left[\ket{\psi,(\alpha \alpha)}+\ket{\psi,(\alpha \beta)}\right]
\end{align}}. 
 
Instead of working with the full wave function, the  AGS formalism introduces transition operators $U_{\beta\alpha}$ describing the transition process from channel $\alpha$ to channel $\beta$~\cite{Mestrom2019}. Using the Faddeev equations~\cite{Faddeev1960, faddeev2013}, the AGS equations can be derived for the transition operator $U$~\cite{alt1967}
\begin{equation} \label{AGS}
    U_{\alpha\beta}= \left(1-\delta_{\alpha\beta}\right)G_0^{-1}+\sum_{\substack{\gamma=1 \\ \gamma \neq \alpha}}^3T_\gamma G_0 U_{\gamma\beta},
\end{equation}
where $G_0$ now represents the free \textit{three-body} Green's function and $T_{\gamma}$ represents the full two-body transition operator, see Eq.~\eqref{LS}, in channel $\gamma$.
Given Eq.~\eqref{AGS}, the three-body transition operators satisfy $U_{\beta \alpha}\ket{\phi,\alpha} = 
V^{\beta}\ket{\psi,\alpha +}$, in direct analogy with the two‑body scattering (cf.\ Sec.~\ref{section: general framework}).
 
In this work we are primarily interested in examining the effective three-body interaction between atoms that are asymptotically free. Such processes are described by the transition matrix $U_{00}$. In the AGS formalism, the equation for $U_{00}$ is coupled to $U_{\alpha 0}$ which describe transitions from three individual atoms to a single atom and a dimer state. Following Ref.~\cite{Mestrom2019}, one can show from Eq.~\eqref{AGS} that these four components of the $U$-matrix are coupled according to
\begin{align}
     U_{00} &= \sum_{\alpha =1}^3 T_\alpha G_0 U_{\alpha 0} \label{eq: AGS U00}\\
    U_{\alpha 0} &= G_0^{-1}+\sum_{\substack{\gamma=1 \\ \gamma \neq \alpha}}^3T_\gamma G_0 U_{\gamma 0} \label{eq: Ualpha0},
\end{align}
where $\alpha=1,2,3$ in the latter expression. 
It is possible to simplify the expression further by introducing a permutation operator acting on the channel indices
\begin{align}
     U_{\alpha 0}= G_0^{-1}+PT_\alpha G_0 U_{\alpha 0}.
\end{align}
Here the permutation operator $P$ consists of a transformation to one channel higher and to one channel lower, $P = P_+ + P_-$, where the plus (minus) sign refers to a (counter-)clockwise-permutation of the three-particle system. With the help of this permutation operator we define~\cite{Mestrom2019Finite}
\begin{equation}\label{def:ubar}
    \bar U_{\alpha 0}=T_\alpha G_0 U_{\alpha 0} (1+P).
\end{equation}
Substituting Eq.~\eqref{def:ubar} into Eqs.~\eqref{eq: AGS U00}-\eqref{eq: Ualpha0} leads to a self-consistent LS equation
\begin{equation}\label{Ubar}
    \bar U_{\alpha 0}=T_\alpha(1+P)+T_\alpha G_0 P \bar U_{\alpha 0}.
\end{equation}
Although in principle one can still work with the original transition operator $U$, the operator $\bar U$ is introduced for convenience, as it can be solved efficiently in momentum space using matrix inversions~\cite{Mestrom2019Finite}. 

\subsection{Separable Model \label{subsec: separable model}}

We solve Eq.~\eqref{Ubar} in momentum space, where the form factors for the two-body interaction are well defined. To that end, we introduce the set of Jacobi momenta $(\mathbf{p}_{\alpha}, \mathbf{q}_{\alpha}$). Here $\mathbf{p}_{\alpha}$ denotes the relative momentum of the interacting pair, while $\mathbf{q}_{\alpha}$ represents the momentum of the spectator particle $\alpha$ with respect to the center of mass of the interacting pair ~\cite{ Glöckle1983, Elster1999, Mestrom2021}. Considering particles of equal mass, the Jacobi momenta are linked to the single-particle momenta $\mathbf{P}_{\alpha}$ as 
\begin{align}
{\bf p}_\alpha &= \frac{1}{2}( {\bf P}_\beta - {\bf P}_\gamma)  \qquad \textrm{and} \\ 
{\bf q}_\alpha &= \frac{2}{3}\left[ {\bf P}_\alpha-\frac{1}{2}( {\bf P}_\beta + {\bf P}_\gamma) \right].
\end{align}
The Jacobi momenta-space states $\ket{\mathbf{p},\mathbf{q}}$ (where the index $\alpha$ was dropped for simplicity) form an orthonormal basis satisfying $\braket{\mathbf{p},\mathbf{q}|\mathbf{p}',\mathbf{q}'} = \delta(\mathbf{p}-\mathbf{p}')\delta(\mathbf{q}-\mathbf{q}')$. These states are eigenfunctions of the free Hamiltonian $H_0 \ket{\mathbf{p},\mathbf{q}} = E_0 \ket{\mathbf{p},\mathbf{q}}$, with eigenvalues $E_0 = p^2 + 3 q^2 /4$. For simplicity we have now set the bare mass to be unity.

When writing Eq.~\eqref{Ubar} in momentum space, it is important that all the sets of the Jacobi coordinates give an equivalent description of the system. In Appendix~\ref{appendix: permutation operators}, a relationship converting $\ket{{\bf p}_1,{\bf q}_1}$ to $\ket{{\bf p}_2,{\bf q}_2}$ or $\ket{{\bf p}_3,{\bf q}_3}$ is presented. The relations can be summarized in the following compact notation
\begin{align}
{\bf p}_i&= \frac{3 }{4}{\bf q}_{i+1}-\frac{1}{2}  {\bf p}_{i+1} = -\frac{3}{4} {\bf q}_{i-1}-\frac{1}{2}{\bf p}_{i-1} \\
{\bf q}_i &= -{\bf p}_{i+1}-\frac{1}{2}{\bf q}_{i+1} = {\bf p}_{i-1}-\frac{1}{2}{\bf q}_{i-1},
\end{align}
where the indices are taken in cyclic order. To keep the notations simple in the following derivations, we introduce a new notation for the momenta according to ${\bf P}^i_0 = \{{\bf p}_{0}, P_+ {\bf p}_{0}, P_- {\bf p}_{0}\} = \{{\bf p}_{0},-\frac{3}{4} {\bf q}_{0}-\frac{1}{2}{\bf p}_{0},\frac{3 }{4}{\bf q}_{0}-\frac{1}{2}  {\bf p}_{0} \}$, where $i=1,2,3$. For $q$ this becomes ${\bf Q}^i_0 = \{{\bf q}_{0}, P_+ {\bf q}_{0}, P_- {\bf q}_{0}\}=  \{{\bf q}_{0},{\bf p}_{0}-\frac{1}{2}{\bf q}_{0},-{\bf p}_{0}-\frac{1}{2}{\bf q}_{0} \}$. 

To write out the free Green's function in momentum space, we use the fact that the Jacobi states are eigenfunctions of the free Hamiltonian $H_0$ with energy: ${\bf p}^2+ 3{\bf q}^2/4$. In this basis, the projection of the free Green's function at a fixed three-body energy, $E$, is given by
\begin{align}\label{Green'sfunction}
    \left\langle  {\bf p}_k,{\bf q}_k \middle| G_0\middle|  {\bf p}_l,{\bf q}_l \right\rangle = \frac{\delta\left({\bf q}_k-{\bf q}_l  \right)\delta\left({\bf p}_k-{\bf p}_l\right)}{E-{\bf p}_l^2 - 3{\bf q}_l^2/4+i\varepsilon} .
\end{align}
Finally, remember that the transition matrix $T_\alpha$ represents the scattering of two particles where the third one remains as a spectator. This is equal to the two-body transition matrix $T(Z_q) = \tau(Z_q) \ket{\xi}\bra{\xi}$ with an energy of $Z_q=E-3q^2/4$, with $\tau(Z)$ given by Eq.~\eqref{T2body}. 

In this basis, $\bar U$ and $U$ can further be expressed as~\cite{Mestrom2021}
\begin{equation}\label{mm}
    \left\langle {\bf p,q}\middle|    U_{00}\middle| {\bf p',q'}\right\rangle=\frac{1}{3} \sum_{\alpha=1}^3  \left\langle {\bf p_\alpha,q_\alpha} \middle|  \bar U_{\alpha 0}\middle| {\bf p',q'}\right\rangle.
\end{equation}

As shown in Appendix \ref{appendix: derivation of eq. u00}, it is possible to solve the self-consistent equation for $\bar U$ in Eq.~\eqref{Ubareq} for each channel and obtain an equation for the transition matrix $U_{00}$. The final result is
\begin{align} 
    &\left\langle {\bf p,q}\middle|U_{00}\middle| {\bf p}_0,{\bf q}_0\right\rangle = -\frac{4\pi}{3} \sum_{\alpha=1}^3 \sum_{i=1}^3 \tau\left(Z_{Q^i_0}\right)\xi\left( P^i_0\right)\frac{\xi(p_\alpha)}{(4\pi)^2} \nonumber\\
    &\times\biggl(-\frac{\delta\left( |{\bf q_\alpha}- {\bf Q}^i_0|\right)}{ q_\alpha^2}  + 4\pi\tau\left(Z_{q_\alpha}\right)   A\left(E,q_\alpha,Q^i_0\right)\biggr), \label{U00 result}
\end{align}
where the sums runs over the different channels and the different Jacobi coordinates. 
Eq.~\eqref{U00 result} consists of two terms. The first term proportional to $\delta\left( |{\bf q_\alpha}- {\bf Q}^i_0|\right)$ comes from the two-body scattering processes resulting from the $T_\alpha(1+P)$. However, this term does not introduce is not related to interacting three-body physics, so it can be disregarded. The second term in Eq.~\eqref{U00 result} defines the three-body scattering amplitude $A\left(E,q_\alpha,Q^i_0\right)$.

Using the explicit forms of the two-body $T$-matrix, Eq.~\eqref{T2body}, and the scattering amplitude, see Appendix \ref{appendix: derivation of eq. u00}, the transition matrix becomes
\begin{align}
       &\left\langle {\bf p,q}\middle|    \mathcal{U}_{00}\middle| {\bf p}_0,{\bf q}_0\right\rangle = - \frac{4\pi}{3} \sum_{\alpha=1}^3\sum_{i=1}^3 \tau\left(Z_{Q^i_0}\right)\xi\left( P^i_0\right)\frac{\xi(p_\alpha) }{(4\pi)^2} \nonumber\\
    &\times \tau\left(Z_{q_\alpha}\right) \biggl[ \int d\hat{\bf Q}^i_0 \frac{\xi\left(\left|{\bf Q}^i_0+\frac{1}{2}{\bf q}_\alpha\right|\right)  \xi\left(\left|{\bf q}_\alpha+\frac{1}{2}{\bf Q}^i_0 \right|\right)}{q_\alpha^2+{\bf q_\alpha}\cdot{\bf Q}^i_0+{Q^i_0}^2-E} \nonumber\\
    &-\frac{4\pi }{2\pi^2} \int dq' q'^2 \int d\hat{\bf q}' \frac{\xi\left(\left|\frac{1}{2}{\bf q'}+{\bf q_\alpha}\right|\right)\xi\left(\left|\frac{1}{2}{\bf q'}+{\bf q_\alpha}\right|\right)}{1/a-\sqrt{-E+3q'^2/4-i\varepsilon}} \nonumber \\
    &\times \frac{ A\left(E,{\bf q'},{\bf Q}^i_0\right)}{q'^2+{\bf q'}\cdot{\bf q_\alpha}+ q_\alpha^{2}-E}  \biggr]. \label{unicorn}
\end{align}
This expression can be simplified further by introducing another version of the scattering amplitude, $\mathcal{A}$, defined as
\begin{align}\label{curly A}
     \left\langle {\bf p,q}\middle| \mathcal{U}_{00}\middle| {\bf p}_0,{\bf q}_0\right\rangle= &-\frac{1}{3}\sum_{\alpha=1}^3 \sum_{i=1}^3 \frac{4}{g_2^2}\tau\left(Z_{Q^i_0}\right) \xi\left( P^i_0\right)\xi(p_\alpha) \nonumber\\
     &\times\tau\left(Z_{q_\alpha}\right) \mathcal{A}\left( E, {\bf q}_\alpha, {\bf Q}^i_0\right).
\end{align}
Comparing this expression to the one in Eq.~\eqref{unicorn}, one can show that the quantity
\begin{equation}
    \mathcal{A}_s\left( E, {\bf q}_\alpha, {\bf Q}^i_0\right) = \int_{-1}^{1} dx \mathcal{A}\left( E, {\bf q}_\alpha, {\bf Q}^i_0\right),
\end{equation}
which corresponds to performing the angular integration over the relative orientation of the momenta ${\bf q}_\alpha$ and ${\bf Q}^i_0$, satisfies the following integral equation
\begin{align}
   &\mathcal{A}_s\left( E, {\bf q}_\alpha, {\bf Q}^i_0\right)= \frac{g_2^2}{4}\nonumber \\
   &\times \tau\left(Z_{q_\alpha}\right)  \int d\hat{\bf Q}^i_0 \frac{\xi\left(\left|{\bf Q}^i_0 + \frac{1}{2}{\bf q}_\alpha\right|\right)  \xi\left(\left|{\bf q}_\alpha+\frac{1}{2}{\bf Q}^i_0 \right|\right)}{q_\alpha^2+{\bf q_\alpha}\cdot{\bf Q}^i_0+{Q^i_0}^2-E} \nonumber\\
    &+\frac{2 }{\pi} \int dq' q'^2 \int d\hat{\bf q}' \frac{\xi\left(\left|\frac{1}{2}{\bf q'}+{\bf q_\alpha}\right|\right)\xi\left(\left|\frac{1}{2}{\bf q'}+{\bf q_\alpha}\right|\right)}{q'^2+{\bf q'}\cdot{\bf q_\alpha}+ q_\alpha^{2}-E} \nonumber \\
    &\times \frac{\mathcal{A}_s\left(E,{\bf q'},{\bf Q}^i_0\right)}{-1/a+\sqrt{-E+3q'^2/4-i\varepsilon}}. \label{unicorn2}
\end{align}

\begin{figure}
    \centering
    \includegraphics[width=0.9\linewidth]{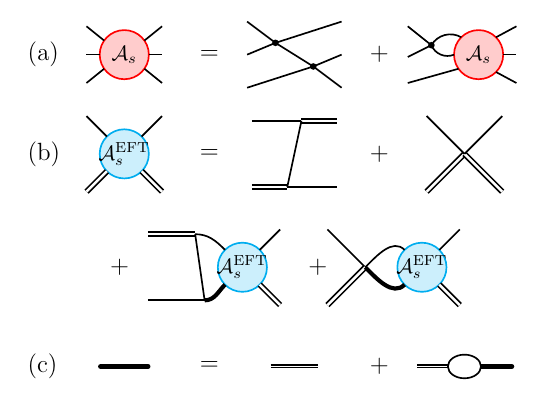}
    \caption{A schematic overview of the scattering amplitudes is shown for the separable model of Eq.~\eqref{AC} in panel (a), and for the EFT model of Eq.~\eqref{As_EFT} in panel (b). Panel (c) displays the corresponding series for the full diatom propagator (thick solid line) given in Eq.~\eqref{eq:DiatomPropagator}. In all panels thin single lines denote single particle propagators, while the double-lines in (b) and (c) denote the free diatom propagator. (Figures (b) and (c) were adapted from \cite{Braaten2006}.) }
    \label{Fig1}
\end{figure}

Eq.~\eqref{unicorn2}, is a self-consistent equation for the scattering amplitude in the zero-angular momentum channel, i.e.\ it is the s-wave scattering amplitude. The remaining angular integrals in Eq.~\eqref{unicorn2} can be compactly noted by defining the function $ H(E,{\bf p},{\bf q})$
\begin{equation}\label{H(p,q)}
    H(E,{\bf p},{\bf q}) = \int_{-1}^{+1} dx \frac{\xi({\bf q}+\frac{{\bf p}}{2})\xi({\bf p}+\frac{{\bf q}}{2})}{p^2+q^2+pqx-E}.
\end{equation}
This integral has known analytical solutions, depending on the relative magnitudes of $p$ and $q$, and can therefore be evaluated straightforwardly once the kinematic regime is specified. 
To bring the result into a simplified form, we first multiply Eq.~(\ref{unicorn2}) by the renormalization factor $Z_D/2 = 64\pi/2ag_2^2$, in direct analogy with the procedure used in the EFT treatment \cite{Braaten2006}. 
We also simplify the momentum variables by making the substitution: ${\bf q}''\rightarrow {\bf q}$, ${\bf Q^i_0}\rightarrow {\bf k}$, and ${\bf q_\alpha}\rightarrow {\bf p}$. With these definitions, the expression for the amplitude takes the following form
\begin{align} \label{Asfinaal}
   \mathcal{A}_s(E,p,k) &=\frac{8\pi}{a}H(E,p,k) \nonumber\\
   &+\frac{2}{\pi} \int dq\,q^2 \frac{H(E,p,q)\,\mathcal{A}_s(E,q,k)}{-1/a+\sqrt{-E+3q^2/4-i\varepsilon}}.
\end{align}
This expression represents the most compact form of the scattering amplitude in the separable potential model. It is an integral equation describing the s-wave three-body scattering amplitude. Eq.~\eqref{Asfinaal} can be pictorially represented by the Feynman diagrams in Fig.~\ref{Fig1}(a). The first term in Eq.~\eqref{Asfinaal}, $H(E,p,k)$, is equivalent to the bare effective three-particle scattering due to two two-particle scattering events. The second term, on the other hand, is denoted by the second term in Fig.~\ref{Fig1}(a), and describes the effect of repeated three-particle scattering.

This method ought to be compared to the EFT approach, which instead uses an explicit three-body contact interaction which need to be regularized.
In Appendix~\ref{subsec: EFT treatment}, we review the literature on EFT methods in evaluating the three-particle scattering amplitude. There we show that an analogous equation can be derived. The role of $H(E,p,q)$ in Eq.~\eqref{Asfinaal} is played by the phenomenological zero-range three-body interaction with strength: $g_3$. A more explicit comparison, see Appendix~\ref{subsec: EFT treatment} shows that:
\begin{align}
   g_3 \underset{\Lambda\rightarrow\infty}{=} \frac{9g_2^2}{2}\int d\hat{\bf k}\frac{\xi\left(\left|{\bf k}+\frac{1}{2}{\bf p}\right|\right)  \xi\left(\left|{\bf p}+\frac{1}{2}{\bf k} \right|\right)-1}{p^2+{\bf p}\cdot{\bf k}+k^{2}-E}.
\end{align}
Thus in our model there is no need to introduce an effective zero-range three-body interaction into the theory.

In order to facilitate a direct comparison with the amplitude from the EFT approach, it is useful to rewrite the product of form factors using the identity: $\xi\left(\left|{\bf p}+\frac{1}{2}{\bf q}\right|\right)  \xi\left(\left|{\bf q}+\frac{1}{2}{\bf p} \right|\right) = 1+ \left[\xi\left(\left|{\bf p}+\frac{1}{2}{\bf q}\right|\right)  \xi\left(\left|{\bf q}+\frac{1}{2}{\bf p} \right|\right)-1\right]$.
Substituting this expression back into Eq.~\eqref{unicorn2} result yields
\begin{widetext} 
\begin{align}\label{AC}
     \mathcal{A}_s&\left( E, p, k\right)=\frac{16\pi}{a} \biggl[\frac{1}{2p k} \ln\left(\frac{p^2+p k+k^{2}-E}{p^2-p k+k^{2}-E}\right) +\frac{1}{2}\int d\hat{\bf k}\frac{\xi\left(\left|{\bf k}+\frac{1}{2}{\bf p}\right|\right)  \xi\left(\left|{\bf p}+\frac{1}{2}{\bf k} \right|\right)-1}{p^2+{\bf p}\cdot{\bf k}+k^{2}-E}  \biggr]\nonumber\\
    &+ \frac{4}{\pi}\int_0^{+\infty} dq\,q^2 \left(\frac{1}{2p q} \ln\left(\frac{p^2+p q+q^2-E} {p^2-p q+q^2-E}\right) +\frac{1}{2} \int d\hat q\frac{\xi\left(\left|\frac{1}{2}{\bf q}+{\bf p}\right|\right)\xi\left(\left|\frac{1}{2}{\bf q}+{\bf p}\right|\right)-1}{q^2+{\bf q}\cdot{\bf p}+ p^{2}-E} \right)\frac{  \mathcal{A}_s\left(E,q,k\right)}{-1/a+\sqrt{-E+3q^2/4-i\varepsilon}} .
\end{align}
\end{widetext}
Eq.~\eqref{AC} highlights the presence of the form factors, in the resulting three-body scattering, and serves as the starting point for our analysis.

\section{The Three-Body Scattering Amplitude \label{section: comparative study}}
We now solve for the three-body scattering amplitude, following Eq.~\eqref{AC} and compare the results from the separable potential and EFT model. \par
In both the separable model and the EFT treatment, we encounter integral equations for the s-wave three-body scattering amplitude similar to Eq.~\eqref{AC}. We solve these integral equations by discretizing momentum space and converting the integral equations into sets of linear equations that can be solved by matrix inversion. The remaining integrals in Eqs.~\eqref{AC} and~\eqref{As_EFT} require a Gaussian quadrature integration scheme \footnote{In this formulation, there are different nodes within the specified integration range, all with their associated weights. For each node, the numerical algorithm computes its contribution to the sum. The used momentum grid is non-uniform and divided into four distinct regions, enabling optimized resolution for both low- and high-momentum scales.}. Below we present the results for the bound states, the inelastic scattering amplitude, and elastic scattering amplitude.

\subsection{Efimov Trimers}
\begin{figure}
\includegraphics[width = 0.45\textwidth]{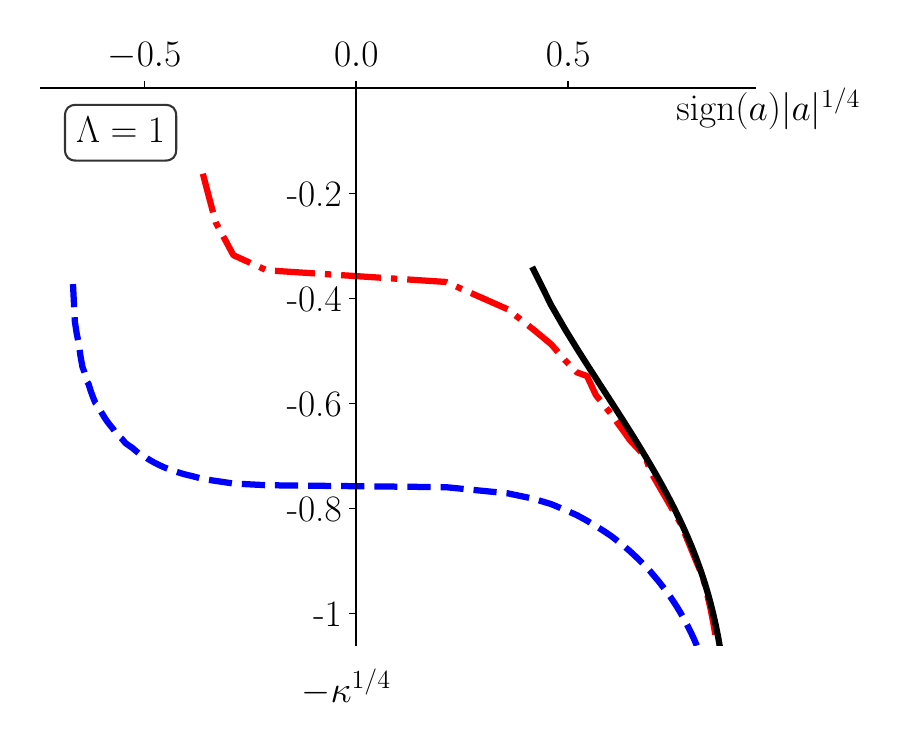}
\caption{Representation of the wavenumbers of the Efimov states as a function of the scattering length $a$ for a cutoff of $\Lambda=1$. The deepest trimers with $\kappa_*^{(0)}$ are displayed here in dashed blue, while the first excited state $\kappa_*^{(1)}$ is plotted here in dotted red. Both these curves end at the black solid line, which represents the dimer binding energy. }
\label{Fig2}
\end{figure}

We benchmark our results by examining the s-wave trimer bound states. The s-wave trimer states are the solutions to the integral equation at negative energies: $E= -\kappa^2$. At resonance, there is a infinite number of bound states with the discrete scaling relation shown in Eq.~\eqref{eq:kappa_ratio}.
In Fig.~\ref{Fig2}, the value of $\kappa$ for the s-wave trimer states are shown in units of $\Lambda$ as a function of the $s$-wave scattering length. The blue curves represent the lowest obtained bound state $\kappa^{(0)}$, while the first excited trimer state is given by the red curve $\kappa^{(1)}$. 
We were also able to obtain $\kappa^{(2)}$ (not shown in Fig.~\ref{Fig2}). 

Qualitatively the trimer spectrum as a function of $1/a$ has the expected characteristic shape obtained from the EFT \cite{Braaten2006}. For small positive scattering lengths, the trimer state merges with the atom-dimer continuum. While for negative scattering lengths, the trimer state will meet the three-atom continuum at a finite scattering length, leading to an enhanced three-body decay \cite{Kraemer2006,pollack2009,Zaccanti2009}. 

At resonance, the ratio of successive trimer energy states $\kappa_*^{(n)}$ is expected to be a constant value, Eq.~\eqref{eq:kappa_ratio}. From our numerical solution we find
\begin{align}
    \kappa^{(1)}/\kappa^{(2)} = 22.6980 \text{     and     } \kappa^{(0)}/\kappa^{(1)} = 24.1884,
\end{align}
in good agreement with the effective field theory result of $22.7$. The departure of $\kappa^{(0)}/\kappa^{(1)}$ from its universal value is due to the fact that the scaling in the Efimov spectrum only holds at low energies, compared to the cutoff. Hence, the state $\kappa^{(0)}$ is most sensitive to finite range effects. \par
\begin{figure}
\includegraphics[width = 0.5\textwidth]{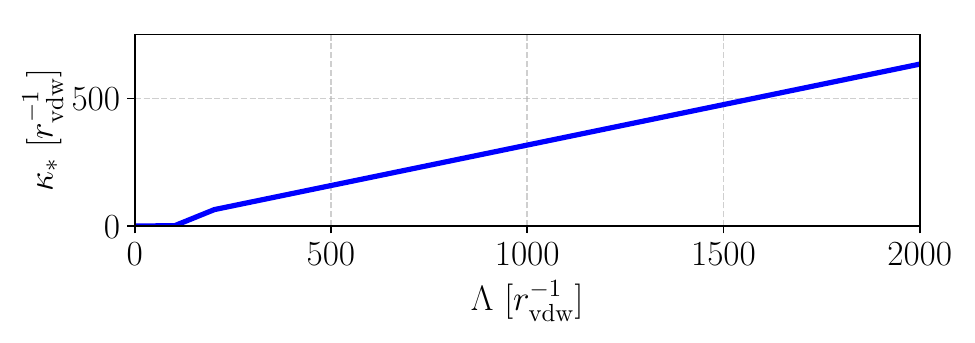}
\caption{Relationship between the Efimov trimer wavenumbers $\kappa_*^{(0)}$ and the cutoff $\Lambda$. All the parameters are expressed in units of the inverse van der Waals length \cite{vandekraats2024}. }
\label{Fig3}
\end{figure}

In addition we consider the $\Lambda$ dependence of the lowest trimer state energy for the separable potential model in Fig.~\ref{Fig3}. For a large range of values we find that $\kappa_*^{(0)}$ depends linearly on $\Lambda$, where $\kappa_*^{(0)} \approx 0.317 \Lambda$. This is expected, as the UV-cutoff is expected to set the deepest trimer, with an energy of order $\Lambda^2$.

\begin{figure}
    \centering
    \includegraphics[width= 0.48\textwidth]{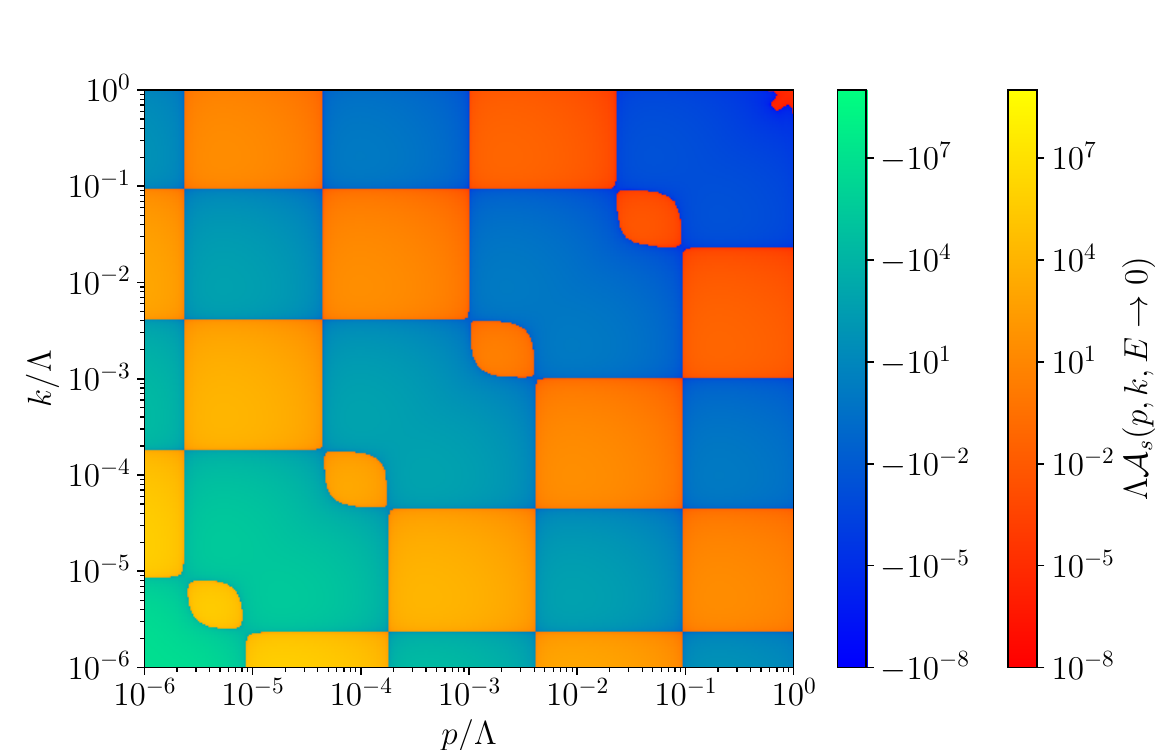}
    \caption{The scattering amplitude $\mathcal{A}_s(p,k,E\rightarrow 0)$ form the separable model presented as a function of the momenta $p/\Lambda$ and $k/\Lambda$. Here, $p$ represents the momentum of the incoming diatom and particle, while $k$ represents the same for the outgoing particle and diatom. This is plotted in the low-energy limit $E \rightarrow 0$. The blue-green color bar denotes regions where the scattering amplitude is negative, while the red-yellow regions denote regions of positive scattering amplitudes.}
    \label{Fig4}
\end{figure}

\subsection{The Scattering Amplitude}
In Fig.~\ref{Fig4}, we present the scattering amplitude $\mathcal{A}_s(p, k, E \rightarrow 0)$ calculated using the separable model as a function of the scaled momenta $p/\Lambda$ and $k/\Lambda$. The amplitude values are shown by the color scale, which is plotted logarithmically. Here we plot positive regions of the scattering amplitude with yellow to red color scheme, while the blue to green color scheme is for the negative values of the scattering amplitude. 
The most prominent feature is the clear log-periodic structure in both arguments, $k$ and $p$ to asymptotically low energies. This is due to the Efimov effect and the lack of a low-energy scale set by the energy or the s-wave scattering length. This structure is particularly evident in the limit $k \to 0$ ($p\to 0$) and finite $p$ ($k$), where we see a regular chess-board pattern emerging. However, when $p\sim k$ this is distorted. We see regimes of small positive scattering amplitude, separated by larger regimes of negative scattering amplitude. Although this structure is still log-periodic, it differs from the inelastic limits, where one momenta tends to zero. This is one of the main observations of this work.

In addition, we find that the s-wave scattering amplitude is symmetric under exchange of the incoming and the outgoing momenta, $p$ and $k$. This is due to time-reversal symmetry that requires:
\begin{equation}
    \mathcal{A}(E,{\bf p},{\bf k}) = \mathcal{A}(E, {\bf -k}, {\bf -p})
\end{equation}
or equivalently: $\mathcal{A}_s(E,p,k) = \mathcal{A}_s(E, k, p)$, as the action of time reversal symmetry does not change the relative angle between ${\bf p}$ and ${\bf k}$.  

\begin{figure}
\includegraphics[width = 0.5\textwidth]{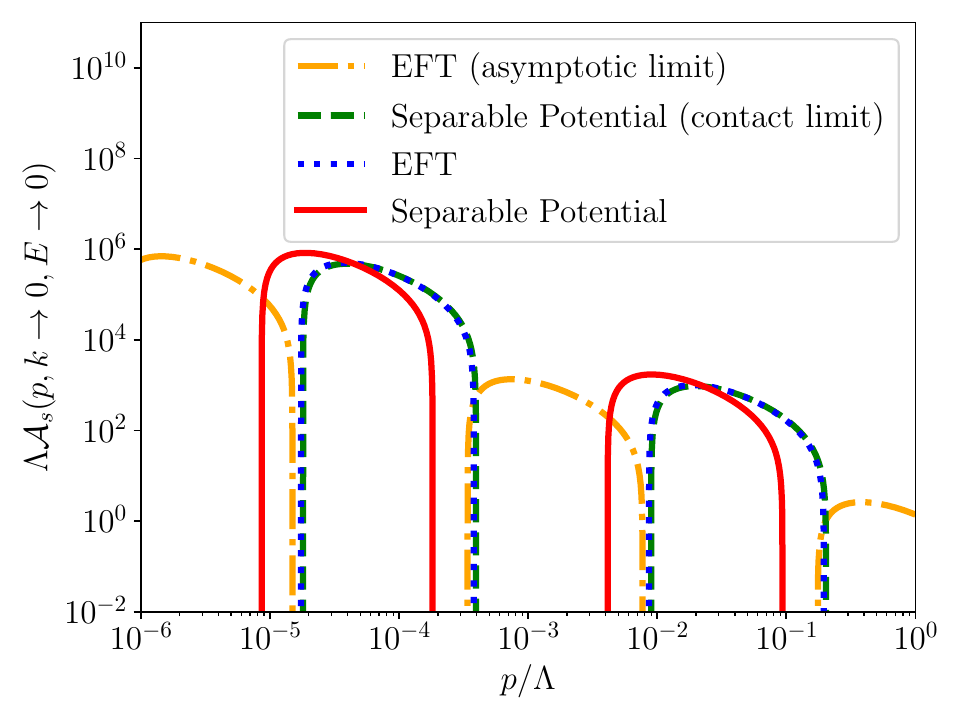}
\caption{The scattering amplitudes $\mathcal{A}_s(p,k\rightarrow 0,E\rightarrow 0)$ presented as a function of the momentum $p$ in units of the cutoff on a log-log scale. Here, $p$ represents the momentum of the incoming diatom and particle, while $k$ represents the same for the outgoing particle and diatom. This is plotted in the low-energy limit $E \rightarrow 0$, and for inelastic scattering where the outgoing particles have small momentum $k \rightarrow 0$. Regions where $A_s(p,k\to 0, E\to 0)<0$ are not shown for simplicity.}
\label{Fig5}
\end{figure}

\begin{figure}
\includegraphics[width = 0.5\textwidth]{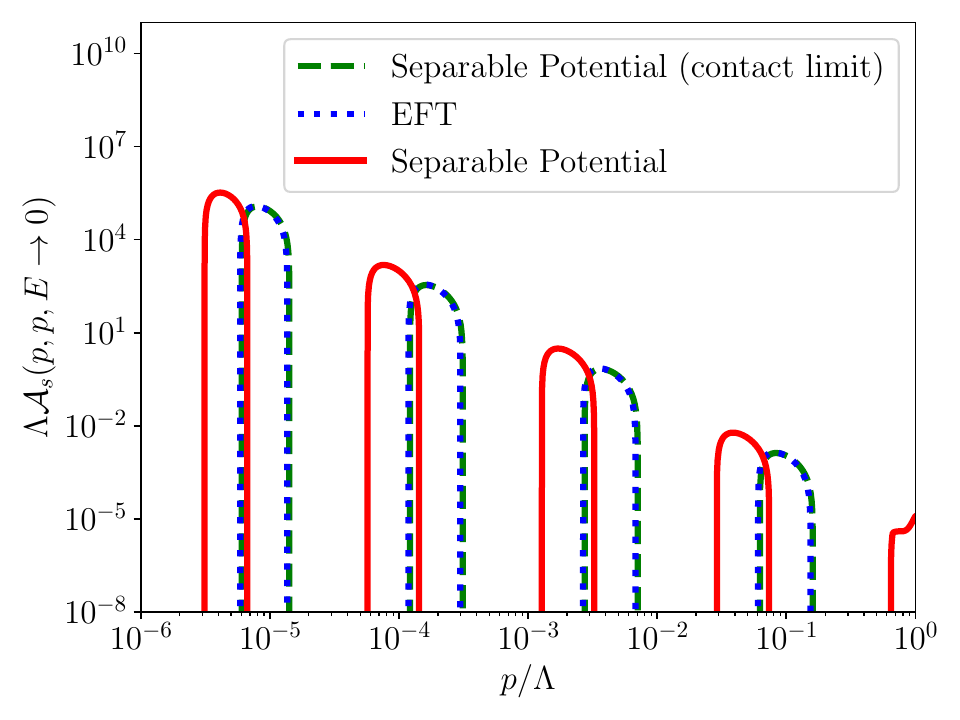}
\caption{The scattering amplitudes $\mathcal{A}_s(p,p,E\rightarrow 0)$ presented as a function of the momentum $p$ in units of the cutoff. Here, $p$ represents the momentum of the incoming diatom and particle, and for the outgoing particle and diatom. This is plotted in the low-energy limit $E \rightarrow 0$. }
\label{Fig6}
\end{figure}

In order to examine the explicit momentum dependence of the scattering amplitude and to compare the separable potential method to the EFT approach, we consider specific cuts of Fig.~\ref{Fig4} in Figs.~\ref{Fig5} and \ref{Fig6}. First, consider the inelastic scattering amplitudes $\mathcal{A}_s(p,k\rightarrow 0,E \rightarrow 0)$ in Fig.~\ref{Fig5}. The red solid line corresponds to the result obtained using the separable interaction model (cf.\ Eq.~\eqref{Asfinaal}) and the green dashed curve represents the contact limit of this expression which is defined when $\Lambda \to \infty$ for fixed $p/\Lambda$. The blue dotted curve shows the EFT prediction for the three-body scattering amplitude (cf.~ Eq.~\eqref{As_EFT}), while the yellow dash-dotted curve represents the analytical solution for the three-body scattering amplitude:
\begin{align}
     \mathcal{A}_s^{EFT}\left( E, p, k\to 0 \right)& \propto \frac{1}{p} \cos\left(s_0 \ln(p/\Lambda^*)\right)
     \label{scal_sol}
\end{align}
which describes the scaling behavior of the s-wave scattering amplitude at large momenta. The quantity $\Lambda^*$ sets the phase of the log-periodic oscillations, and needs to be set by comparing the EFT method to experiments, or in this case, by comparing to our separable potential method.

Notably, all computed scattering amplitudes exhibit the expected power-law decay $p^{-1}$ consistent with the analytic large-momentum behavior of $\mathcal{A}_s(E,p \rightarrow \infty,k)$ given in Eq.~\eqref{scal_sol}. Physically, the decay of the scattering amplitudes with momentum reflects that atom–diatom scattering at large relative momentum $p$ is less sensitive to the resonant atom-dimer state in the unitary limit. The scattering amplitude also exhibits clear log-periodic oscillations consistent with the Efimov effect~\cite{Moroz2009, Hammer2010, Kievsky2021}, independent of the method used. The period of the oscillations is equivalent for all models and is set by the parameter $s_0$.

The main difference in methods is a phase shift in the oscillations. This phase shift between the EFT and the separable potential method is a consequence of the finite interaction range.
In the contact limit, where no finite range effects are present, we find perfect agreement between the separable potential method and the full EFT solution. The agreement between these two approaches can be clearly seen by noting that the contact limit in Eq.~\eqref{Asfinaal} reduces the function $H(E,p,q)$ to the same logarithmic expression that appears in the EFT amplitude. Consequently, the contact-limit result of our separable model (green dashed curve) coincides with the EFT prediction (blue dotted curve). However, we see that the asymptotic solution of the EFT, Eq.~\eqref{scal_sol}, does not have the correct phase shift, compared to the full solution. This phase shift has important consequences for the scattering properties at low-momenta, especially in the low-energy limit, as the phase of the log-periodic oscillations set the strength of the scattering amplitude. Also provided we set $c\approx 1.31$, we find a perfect overlap between the general separable potential solution and the EFT, which disagrees with the reported value of $c=2.62$ \cite{Braaten2006}.

We now turn to the analysis of elastic atom–diatom scattering, where $k = p$. In contrast to the inelastic case considered above ($k = 0$ and finite $p/\Lambda$) there is, to the best of our knowledge, no closed-form expression for the three-body elastic scattering amplitude $\mathcal{A}_s(p,p,0)$. The results of our analysis are shown in Fig.~\ref{Fig6}, where we show the positive part of the scattering amplitude as a function of $p/\Lambda$ on a log-log scale.
As in the inelastic case, we find that the contact limit of the separable interaction model (green dashed curves) faithfully reproduces the EFT prediction (blue dotted curves).

The elastic s-wave scattering amplitude also exhibits log-periodic behavior, with a decay in the amplitude of the oscillations at large momenta. However, there are two noticeable differences. The first is that the periodicity of the log-periodic oscillations in $\mathcal{A}_s(p,p,0)$ is about twice as fast as in the inelastic case. This can be clearly seen in Fig.~\ref{Fig4}, where the elastic scattering amplitude clearly exhibits a different periodicity than in the inelastic case. Secondly, the amplitude now exhibits a $p^{-2}$ decay, rather than the $p^{-1}$ behavior observed for the inelastic scattering amplitude. Such features are well described by a scaling solution of the form:
\begin{align}
     \mathcal{A}_s^{EFT}\left( E, p, k\to 0 \right)& \propto \frac{1}{p^2} \cos\left(2s_0 \ln(p/\Lambda^*)\right)
     \label{scal_sol}.
\end{align}
We posit that such scaling forms exist for all rays in $p-k$ space with fixed $p/k$ with variable periodicities, as evidenced from Fig.~\ref{Fig4}.

\section{Conclusion \label{section: conclusion}}
In this work we have re-examined the three-body scattering of identical bosons with resonantly large, finite ranged, two-body interactions, and compared it to the zero-range effective field theory method. To accomplish this, we reformulated the STM equations using a separable potential. This involved applying a Faddeev decomposition, which assumed that the three particles interact only through pairwise forces. From this framework, we derived an integral equation for the three-body s-wave scattering amplitude, denoted as $\mathcal{A}_s(p,k,E)$. \par

The three-body scattering amplitude with separable interactions reproduces the behavior expected from EFT without requiring regularization or renormalization. The resulting amplitudes display a logarithmic-periodic scaling and are symmetric in the incoming and outgoing momenta, $p$ and $k$.  
For inelastic scattering, both models yield comparable results: the same oscillation width, period and an overall $1/p$ decay. However, the models differ in the presence of a phase shift in the log-periodic oscillations, which is relevant in predicting the low-energy behavior of the s-wave scattering amplitude. We find that the period of the oscillations and the decay of the envelope of the scattering amplitude depends on the value of $p$ and $k$. For elastic scattering, the oscillation period is reduced by half compared to the inelastic case, and all amplitudes decay as $1/p^2$.

Nevertheless, there are several open questions that could be further investigated using this method. One promising direction would be to extend our framework to include the effective range parameter, which has been shown to be important in describing deep Trimer states and three-body recombination rates \cite{Dyke2013,Srensen2013}. While this term can be readily incorporated into the analytical expression, it generally leads to perturbative effects in interacting s-wave bosonic systems. However, the effective range is important in describing quantum gases with p-wave gases \cite{jona2008,Yu2015,Zhu2022}. Our analysis can readily explore the role of the effective range on interesting few-body structures, like the super-Efimov effect in p-wave Fermi gases~\cite{Nishida2013}. 
Additionally, our analysis was restricted to the unitary limit, corresponding to a single value of the scattering length. The log-periodic oscillations in the scattering amplitude will naturally be cut off at low energies by the presence of a finite scattering length, or finite energy. The analytical derivations of the scattering amplitude with the separable potential remain valid for arbitrary scattering lengths, allowing our theory to be extended beyond the strongly interacting case to weakly interacting regimes, where EFT formulations are also available~\cite{Lee1957,Salasnich1998,Petrov2015}.

\begin{acknowledgments}
We acknowledge financial support from the Research Foundation Flanders (FWO) project nos. G0AIY25N, G0A9F25N and G060820N. D. Ahmed-Braun acknowledges funding from the Research Foundation-Flanders via a postdoctoral fellowship (Grant No. 1222425N). Furthermore, we thank Servaas Kokkelmans and Jasper van de Kraats for fruitful discussions. 
\end{acknowledgments}

\bibliography{bibliography}

\appendix
\begin{widetext}
\section{Permutation Operators \label{appendix: permutation operators}}
To understand how the permutations between the different channel descriptions of the Jacobi coordinates work, it is easier to work in the basis of the single-particle momenta. For example, in channel 1, the Jacobi coordinates are defined as ${\bf p}_1= \left( {\bf P}_2 - {\bf P}_3\right)/2$, and ${\bf q}_1= 2\left[ {\bf P}_1-\left( {\bf P}_2 + {\bf P}_3\right)/2 \right]/3$. From these relations, the inverse transformation, expressing the single-particle momenta in terms of the Jacobi coordinates, can be obtained~\cite{Elster1999}. When we impose the center of mass condition for the three-particle system, ${\bf P}_1+{\bf P}_2+{\bf P}_3=0$, the single-particle momenta become
\begin{align}
    {\bf P}_1&={\bf q}_1 \\
    {\bf P}_2&= {\bf p}_1-\frac{1}{2}{\bf q}_1 \\
    {\bf P}_3&= -\frac{1}{2}{\bf q}_1-{\bf p}_1.
\end{align}
For channels $\alpha=2,3$, the same procedure can be repeated analogously. The results are summarized in Table \ref{tablejac}. The three sets of coordinates must be mutually equivalent, meaning that any set can be transformed into another. To illustrate this, consider the transform from channel 1 to channel 2. Using the relations for identical particles in the table, one finds the mapping ${\bf q}_1\rightarrow -{\bf q}_2/2-{\bf p}_2$. This mapping ensures that the momentum ${\bf P}_1$ already match.

Requiring that the remaining components ${\bf P}_2$ and ${\bf P}_3$ also coincide across coordinate sets further yields the transformation for the second Jacobi momentum: ${\bf p}_1\rightarrow -{\bf p}_2/2+3{\bf q}_2/4$. 
A similar procedure applies to the transformation from channel 1 to channel 3, leading to the relations ${\bf q}_1\rightarrow {\bf p}_3-{\bf q}_3/2$ and ${\bf p}_1\rightarrow -3 {\bf q}_3/4-{\bf p}_3/2$. \\
From these transformation rules, follows the operation of the permutation operator in the Jacobi momentum basis:
\begin{align}
   \left\langle {\bf p}_j,{\bf q}_j \middle| P\middle|{\bf p}_i,{\bf q}_i\right\rangle &= \left\langle {\bf p}_j,{\bf q}_j \middle|( P_+ + P_-)\middle|{\bf p}_i,{\bf q}_i\right\rangle\nonumber \\
   &= \left\langle {\bf p}_j,{\bf q}_j \middle| {\bf p}_{i+1},{\bf q}_{i+1}\right\rangle + \left\langle {\bf p}_j,{\bf q}_j \middle| {\bf p}_{i-1},{\bf q}_{i-1}\right\rangle\nonumber \\
   &=\braket{ {\bf p}_j,{\bf q}_j |-\frac{3}{4} {\bf q}_{i}-\frac{1}{2}{\bf p}_{i},{\bf p}_{i}-\frac{1}{2}{\bf q}_{i} }+ \braket{ {\bf p}_j,{\bf q}_j | \frac{3 }{4}{\bf q}_{i}-\frac{1}{2}  {\bf p}_{i}, -{\bf p}_{i}-\frac{1}{2}{\bf q}_{i}}. \label{P}
\end{align}
Remembering that the Jacobi coordinates form an orthogonal basis, this can be further simplified to 
\begin{align}
    \left\langle {\bf p}_j,{\bf q}_j \middle| P\middle|{\bf p}_i,{\bf q}_i\right\rangle &= \delta\left({\bf p}_j+\frac{3}{4} {\bf q}_{i}+\frac{1}{2}{\bf p}_{i} \right) \delta\left({\bf q}_j-{\bf p}_{i}+\frac{1}{2}{\bf q}_{i} \right) + \delta\left({\bf p}_j-\frac{3 }{4}{\bf q}_{i}+\frac{1}{2}  {\bf p}_{i} \right) \delta\left({\bf q}_j+{\bf p}_{i}+\frac{1}{2}{\bf q}_{i}\right) \\
   &= \delta\left({\bf p}_j +\frac{1}{2}{\bf q}_j+{\bf q}_{i}  \right) \delta\left({\bf q}_j-{\bf p}_{i}+\frac{1}{2}{\bf q}_{i} \right) + \delta\left({\bf p}_j-\frac{1 }{2}{\bf q}_{j}- {\bf q}_{i} \right) \delta\left({\bf q}_j+{\bf p}_{i}+\frac{1}{2}{\bf q}_{i}\right).
\end{align}

\begin{table}
\centering
\begin{tabular}{|c|c|c|c|}
\hline
Jacobi coordinates & ${\bf P}_1$ & ${\bf P}_2$ & ${\bf P}_3$ \\ \hline
$\ket{{\bf p}_1,{\bf q}_1}$ & ${\bf q}_1$     &${\bf p}_1-\frac{1}{2}{\bf q}_1 $   &  $-\frac{1}{2}{\bf q}_1-{\bf p}_1$  \\ \hline
 $\ket{{\bf p}_2,{\bf q}_2}$   &$-\frac{1}{2}{\bf q}_2-{\bf p}_2$     &   ${\bf q}_2$ & ${\bf p}_2-\frac{1}{2}{\bf q}_2 $    \\ \hline
       $\ket{{\bf p}_3,{\bf q}_3}$            & ${\bf p}_3-\frac{1}{2}{\bf q}_3 $     &$-\frac{1}{2}{\bf q}_3-{\bf p}_3$    & ${\bf q}_3$    \\ \hline
\end{tabular}
\caption{Single-particle momenta in the center of mass of the three-particle system as a function of the Jacobi coordinates.} 
\label{tablejac}
\end{table}

\section{Derivation of Eq.~(\ref{U00 result})\label{appendix: derivation of eq. u00}}
We aim to find an analytical expression for the transition operator $U_{00}$. Therefore, we will start from Eq.~(\ref{mm}) and fill in the expansion for $\bar U$:
\begin{align}\label{Ubareq}
 \left\langle {\bf p,q}\middle| U_{00}\middle|  {\bf p}_0,{\bf q}_0\right\rangle &= \frac{1}{3} \sum_{\alpha=1}^3  \left\langle {\bf p_\alpha,q_\alpha} \middle|  \bar U_{\alpha 0}\middle|  {\bf p}_0,{\bf q}_0\right\rangle \nonumber \\
  & = \frac{1}{3} \sum_{\alpha=1}^3 \left\langle  {\bf p}_j,{\bf q}_j \middle|T_\alpha(1+P)+T_\alpha G_0 PT_\alpha(1+P) + T_\alpha G_0 PT_\alpha G_0 P T_\alpha(1+P)+\cdots \middle| {\bf p}_0,{\bf q}_0 \right\rangle .
\end{align}
The simplest approach is to analyze each term individually, and identify how the initial terms reappear in the subsequent ones. For the first term, we need to solve for 
\begin{align}\label{eq: first term}
 \frac{1}{3} \sum_{\alpha=1}^3\left(  \left\langle {\bf p_\alpha,q_\alpha} \middle|  T_\alpha(Z)\middle|  {\bf p}_0,{\bf q}_0\right\rangle +  \left\langle {\bf p_\alpha,q_\alpha} \middle|  T_\alpha(Z)P\middle|  {\bf p}_0,{\bf q}_0\right\rangle \right).
\end{align}
Using the notations where 
${\bf P}^i_0 = \{{\bf p}_{0}, P_+ {\bf p}_{0}, P_- {\bf p}_{0}\} = \{{\bf p}_{0},-\frac{3}{4} {\bf q}_{0}-\frac{1}{2}{\bf p}_{0},\frac{3 }{4}{\bf q}_{0}-\frac{1}{2}  {\bf p}_{0} \}$, and ${\bf Q}^i_0 = \{{\bf q}_{0}, P_+ {\bf q}_{0}, P_- {\bf q}_{0}\}=  \{{\bf q}_{0},{\bf p}_{0}-\frac{1}{2}{\bf q}_{0},-{\bf p}_{0}-\frac{1}{2}{\bf q}_{0} \}$, this can be further written as
\begin{align}
    &=\frac{1}{3} \sum_{\alpha=1}^3\Bigl(  \delta\left( {\bf q_\alpha}- {\bf Q}^1_0\right) \left\langle {\bf p_\alpha} \middle|  T_\alpha(Z)\middle| {\bf P}^1_0\right\rangle +  \delta\left( {\bf q_\alpha}- {\bf Q}^2_0\right)\left\langle {\bf p_\alpha} \middle|  T_\alpha(Z)\middle|  {\bf P}^2_0\right\rangle + \delta\left( {\bf q_\alpha}- {\bf Q}^3_0\right) \left\langle {\bf p_\alpha} \middle|  T_\alpha(Z)\middle|  {\bf P}^3_0\right\rangle\Bigr).
\end{align}
In this expression, a summation over the different channels can be recognized. Writing the $T$-matrix in its separable form results in
\begin{align}\label{U00term1}
\frac{1}{3} \sum_{\alpha=1}^3  \left\langle {\bf p_\alpha,q_\alpha} \middle|  T_\alpha(Z)(1+P)\middle| {\bf p}_0,{\bf q}_0\right\rangle &= \frac{1}{3} \sum_{\alpha=1}^3\sum_{i=1}^3 \delta( {\bf q_\alpha}- {\bf Q}^i_0) \left\langle {\bf p_\alpha} \middle|  T_\alpha(Z)\middle|  {\bf P}^i_0\right\rangle \nonumber \\
& =  \frac{1}{3} \sum_{\alpha=1}^3\sum_{i=1}^3 \tau(Z_{Q^i_0}) \xi( p_\alpha) \xi( P^i_0) \frac{\delta\left( |{\bf q_\alpha}- {\bf Q}^i_0|\right)}{4\pi q_\alpha^2}.
\end{align}
For the second term in the expansion of $\bar U$, a similar analysis can be performed. The term that needs to be solved is given by
\begin{align}
    \frac{1}{3} \sum_{\alpha=1}^3  \left\langle {\bf p_\alpha,q_\alpha} \middle| T_\alpha(Z)G_0P T_\alpha(Z)(1+P)\middle| {\bf p}_0,{\bf q}_0\right\rangle.
\end{align}
The Jacobi coordinates form an orthogonal basis where multiple identity relationships can be inserted between the operators,
\begin{align}
    \left\langle {\bf p,q}\middle|    U_{00}^{(term \,2)}\middle| {\bf p}_0,{\bf q}_0\right\rangle  =&\frac{1}{3} \sum_{\alpha=1}^3 \int d{\bf q'}d{\bf p'}d{\bf q''}d{\bf p''} \left\langle {\bf p_\alpha,q_\alpha} \middle| T_\alpha(Z)\middle| {\bf p'},{\bf q'}\right\rangle \left\langle {\bf p',q'} \middle| G_0P \middle| {\bf p''},{\bf q''}\right\rangle \nonumber\\
    &\times\left\langle {\bf p'',q''} \middle| T_\alpha(Z)(1+P)\middle| {\bf p}_0,{\bf q}_0\right\rangle. \label{nn}
\end{align}
The last braket in this expression is again equal to the first term in the expansion of the $\bar U$-matrix, which can be seen from Eq.~(\ref{U00term1}). Substituting the result in this equation, and working out the $T$-matrix and the permutation operator gives
\begin{align}
    =&\frac{1}{3} \sum_{\alpha=1}^3 \int d{\bf q'}d{\bf p'}d{\bf q''}d{\bf p''} \delta\left({\bf q_\alpha}-{\bf q'} \right)\tau(Z_{q_\alpha})\xi(p_\alpha)\xi(p')\times \sum_{i=1}^3 \tau(Z_{Q^i_0}) \xi( p'') \xi( P^i_0) \frac{\delta\left( |{\bf q''}- {\bf Q}^i_0|\right)}{4\pi q''^2}\nonumber\\
   &\times\left[ \left\langle {\bf p',q'} \middle| G_0 \middle| -\frac{3}{4} {\bf q''}-\frac{1}{2}{\bf p''},{\bf p''}-\frac{1}{2}{\bf q''}\right\rangle +
   \left\langle {\bf p',q'} \middle| G_0 \middle| \frac{3 }{4}{\bf q''}-\frac{1}{2}  {\bf p''} ,-{\bf p''}-\frac{1}{2}{\bf q''}\right\rangle\right].
\end{align}
The Green's function was defined in Eq.~(\ref{Green'sfunction}), which can be used here to find
\begin{align}
     &= \frac{1}{3} \sum_{\alpha=1}^3 \int d{\bf q'}d{\bf p'}d{\bf q''}d{\bf p''} \delta\left({\bf q_\alpha}-{\bf q'} \right)\tau(Z_{q_\alpha})\xi(p_\alpha)\xi(p')\times \sum_{i=1}^3 \tau(Z_{Q^i_0}) \xi( p'') \xi( P^i_0) \frac{\delta\left( |{\bf q''}- {\bf Q}^i_0|\right)}{4\pi q''^2}\nonumber\\
   &\times\left[ \frac{\delta\left({\bf p'}+\frac{3}{4} {\bf q''}+\frac{1}{2}{\bf p''}\right)\,\delta\left({\bf q'}-{\bf p''}+\frac{1}{2}{\bf q''}\right)}{E-\left({\bf p''}^2+\frac{3}{4}{\bf q''}^2\right)} +\frac{\delta\left({\bf p'} -\frac{3 }{4}{\bf q''}+\frac{1}{2}  {\bf p''}\right)\,\delta\left({\bf q'}+{\bf p''}+\frac{1}{2}{\bf q''}\right)}{E-\left({\bf p''}^2+\frac{3}{4}{\bf q''}^2\right)}\right]. \label{oo}
\end{align}
These two terms have a same denominator, but have different numerators with distinct delta functions. However, these two terms will still give a similar result in the end. This is the result of the  choice of the separable potential. Since the form factors that belong to the potential only depend on the magnitude of the momenta (not the directions), working out the delta functions in this equation will lead to form factors with the same arguments for both terms. Because of this, we will only work with one of these Green's function, but with an extra factor of  $\times 2$. We continue with the first term, where working out the delta functions leads to 
\begin{align}
    = \frac{1}{3} &\sum_{\alpha=1}^3\sum_{i=1}^3 \int d{\bf q''}d{\bf p''} \delta\left({\bf q_\alpha}-{\bf p''}+\frac{1}{2}{\bf q''} \right)\tau(Z_{q_\alpha})\xi(p_\alpha)\xi\left(\left|-\frac{3}{4} {\bf q''}-\frac{1}{2}{\bf p''}\right|\right)  \tau(Z_{Q^i_0}) \xi( p'') \xi( P^i_0) \nonumber\\
   &\times \frac{\delta\left( |{\bf q''}- {\bf Q}^i_0|\right)}{4\pi q''^2}\frac{2}{E-\left({\bf p''}^2+\frac{3}{4}{\bf q''}^2\right)}.\label{ppp}
\end{align}
The last remaining integrals can be solved with the delta functions and the use of spherical coordinates. The final result can be expressed as
\begin{align}
     = \frac{1}{3} &\sum_{\alpha=1}^3\sum_{i=1}^3 \int d\hat{\bf Q}^i_0\tau\left(Z_{q_\alpha}\right)\xi(p_\alpha)\xi\left(\left|{\bf Q}^i_0+\frac{1}{2}{\bf q}_\alpha\right|\right)  \tau\left(Z_{Q^i_0}\right) \xi\left(\left|{\bf q}_\alpha+\frac{1}{2}{\bf Q}^i_0 \right|\right) \xi\left( P^i_0\right) \frac{2\pi}{4\pi
     }\frac{2}{E-\left(q_\alpha^2+{\bf q_\alpha}\cdot{\bf Q}^i_0+{Q^i_0}^2\right)} ,
\end{align}
where we use the notation $d\hat{\bf Q}^i_0=d(\cos\theta)$. This can be written in more compact form as
\begin{align}
    \frac{1}{3} \sum_{\alpha=1}^3  \left\langle {\bf p_\alpha,q_\alpha} \middle| T_\alpha(Z)G_0P T_\alpha(Z)(1+P)\middle| {\bf p}_0,{\bf q}_0\right\rangle=-\frac{1}{6} \sum_{\alpha=1}^3\sum_{i=1}^3 \tau\left(Z_{q_\alpha}\right)\xi(p_\alpha)\tau\left(Z_{Q^i_0}\right)  \xi\left( P^i_0\right)2U_1\left(E,{\bf q_\alpha},{\bf Q}^i_0\right),
\end{align}
where the angular integral is defined according to 
\begin{align} 
    U_1\left(E,{\bf q_\alpha},{\bf Q}^i_0\right) = \int d\hat{\bf Q}^i_0\, \xi\left(\left|{\bf Q}^i_0+\frac{1}{2}{\bf q}_\alpha\right|\right)  \xi\left(\left|{\bf q}_\alpha+\frac{1}{2}{\bf Q}^i_0 \right|\right)\frac{1}{q_\alpha^2+{\bf q_\alpha}\cdot{\bf Q}^i_0+{Q^i_0}^2-E}.
\end{align}
Finally, the third term that we want to determine is 
\begin{align}
    \frac{1}{3} \sum_{\alpha=1}^3  \left\langle {\bf p_\alpha,q_\alpha} \middle| T_\alpha(Z)G_0PT_\alpha(Z)G_0P T_\alpha(Z)(1+P)\middle| {\bf p}_0,{\bf q}_0\right\rangle.
\end{align}
Inserting the necessary identity relationships leads to 
\begin{align}
    &=\frac{1}{3} \sum_{\alpha=1}^3 \int d{\bf q'}d{\bf p'}d{\bf q''}d{\bf p''} d{\bf q'''}d{\bf p'''}d{\bf q''''}d{\bf p''''} \delta\left({\bf q_\alpha}-{\bf q'}\right) \tau(Z_{q_\alpha})\xi(p_\alpha)\xi(p') \left\langle {\bf p',q'} \middle| G_0 P\middle|{\bf p'',q''} \right\rangle\nonumber \\
    &\times \delta\left({\bf q''}-{\bf q'''}\right) \tau(Z_{q''})\xi(p'')\xi(p''') \left\langle {\bf p''',q'''} \middle| G_0 P\middle|{\bf p'''',q''''} \right\rangle \sum_{i=1}^3 \tau(Z_{Q^i_0}) \xi( p'''') \xi( P^i_0) \frac{\delta\left( |{\bf q''''}- {\bf Q}^i_0|\right)}{4\pi q''''^2}.
\end{align}
Working this out in exactly the same way as before, the expression reduces to 
\begin{align}
    &\frac{1}{3} \sum_{\alpha=1}^3 \left\langle {\bf p_\alpha,q_\alpha} \middle| T_\alpha(Z)G_0PT_\alpha(Z)G_0P T_\alpha(Z)(1+P)\middle| {\bf p}_0,{\bf q}_0\right\rangle \\
    &=\frac{8\pi}{6}\sum_{\alpha=1}^3\sum_{i=1}^3\tau(Z_{Q^i_0})\xi( P^i_0)\xi(p_\alpha) \tau(Z_{q_\alpha})\int dq' q'^2  U_1(E,{\bf q}_\alpha,{\bf q'})\tau(Z_{q'})U_1(E,{\bf q'},{\bf Q}^i_0) .
\end{align}
Now putting all three terms back together allows us to find the expression for the transition operator:
\begin{align}
     \left\langle {\bf p,q}\middle|    U_{00}\middle| {\bf p}_0,{\bf q}_0\right\rangle = -\frac{4\pi}{3} &\sum_{\alpha=1}^3 \sum_{i=1}^3 \tau\left(Z_{Q^i_0}\right) \xi\left(P^i_0\right) \frac{\xi(p_\alpha)}{(4\pi)^2} \biggl(\frac{-\delta\left( |{\bf q_\alpha}- {\bf Q}^i_0|\right)}{ q_\alpha^2}  + 4\pi \tau\left(Z_{q_\alpha}\right) \biggl[U_1\left(E,{\bf q_\alpha},{\bf Q}^i_0\right)\nonumber \\
     &-4\pi\int dq' q'^2 \,U_1\left(E,{\bf q}_\alpha,{\bf q'}\right)\tau(Z_{q'})U_1\left(E,{\bf q'},{\bf Q}^i_0\right)+\cdots\biggr]\biggr).
\end{align}
Here a self-repeating expression can be recognized, which can be labeled as the scattering amplitude
\begin{equation}\label{A}
    A\left(E,q_\alpha,Q^i_0\right) =  U_1\left(E,{\bf q_\alpha},{\bf Q}^i_0\right) - 4 \pi\int dq' q'^2\,U_1\left(E,{\bf q}_\alpha,{\bf q'}\right)\tau(Z_{q'})A\left(E,{\bf q'},{\bf Q}^i_0\right).
\end{equation}
The final expression we find is equal to the result in Eq. (\ref{U00 result}):
\begin{align}
    \left\langle {\bf p,q}\middle| U_{00}\middle| {\bf p}_0,{\bf q}_0\right\rangle = -\frac{4\pi}{3} \sum_{\alpha=1}^3 \sum_{i=1}^3 \tau\left(Z_{Q^i_0}\right) \xi\left( P^i_0\right) \frac{\xi(p_\alpha) }{(4\pi)^2} \biggl(\frac{-\delta\left( |{\bf q_\alpha}- {\bf Q}^i_0|\right)}{ q_\alpha^2}  + 4\pi \tau\left(Z_{q_\alpha}\right)   A\left(E,q_\alpha,Q^i_0\right)\biggr).
\end{align}

\section{EFT Treatment of Three-Body Scattering}
\label{subsec: EFT treatment}

We compare our results using Eq.~\eqref{AC} to the effective field theory approach. For the sake of completeness, we sketch the derivation here, while detailed discussions can be found in the literature \cite{bedaque1999,Bedaque1999renormalization, Braaten2001,Braaten2003,Braaten2006}. The starting point is the following Lagrangian density:
\begin{equation}
    \mathcal{L} = \psi^{\dagger}\left(i\partial_t + \frac{1}{2}\nabla^2 \right) \psi + \frac{g_2(\Lambda)}{4}  |\psi|^4 + \frac{g_3(\Lambda)}{36} |\psi|^6.
    \label{eq:eft_lagrangian}
\end{equation}
where $\psi^{(\dagger)}(x)$ is the annihilation (creation operator of a bosonic particle at position $x$, while $g_2(\Lambda)$ and $g_3(\Lambda)$ are the coupling constants for the zero-range two- and three-body interactions.
Eq.~\eqref{eq:eft_lagrangian} is the simplest zero-range model that qualitatively describes the low-energy physics of interacting bosons with both two-body and three-body interactions. However, because this model is zero-range, two- and three-particle scattering suffer from UV divergences. Thus it is necessary to find appropriate expressions for $g_2(\Lambda)$ and $g_3(\Lambda)$ that remove any dependence of the low-energy physics on the UV scale, $\Lambda$. This can be done by examining the two- and three particle scattering amplitudes, similar to the discussion in Secs.~\ref{section: general framework} and~\ref{subsection: AGS formalism}.

In order to derive an expression equivalent to Eq.~\eqref{AC}, one follows a similar approach as in Sec.~\ref{subsec: separable model} and first solve for the two-body scattering amplitude. This can be done by using a Hubbard Stratonovich transformation which introduces a diatom field, $d$. The Lagrangian for the system is then:
\begin{align}
    \mathcal{L} &= \psi^{\dagger}\left(i\partial_t + \frac{1}{2}\nabla^2 \right) \psi + \frac{g_2}{4}  d^{\dagger}d - \frac{g_2}{4} \left(d^{\dagger}\psi\psi + \psi^{\dagger}\psi^{\dagger} d\right)- \frac{g_3}{36}d^{\dagger}d \psi^{\dagger}\psi .
    \label{eq:lagrangian_Diatom}
\end{align}
One can show by integrating out the field $d$, that Eq.~\eqref{eq:lagrangian_Diatom} is equivalent to Eq.~\eqref{eq:eft_lagrangian} by only keeping the terms up to cubic order in fields $\psi\psi^\dagger$. Physically, Eq.~\eqref{eq:lagrangian_Diatom} resembles a two-channel model, where two bosons can form a diatomic molecule, $d$, that has no dynamics. In the presence of the vacuum, it is possible to obtain an exact expression for the diatom propagator, $D(Z)$, with off-shell energy $Z$, accounting for quantum fluctuations arising from the spontaneous creation and annihilation of two bosons. The expression for the full diatom propagator $D$ is presented in Fig.~\ref{Fig1}c). Exploiting the analogy with scattering in the Feshbach formalism, we derive the exact form of this dressed propagator in App.~\ref{sec:The two-channel model}, finding that 
\begin{align}
\label{eq:DiatomPropagator}
D(Z) = \frac{4}{g_2}\left[1+\frac{g_2}{4\pi^2}(\Lambda-\frac{\pi}{2}\sqrt{-Z-i\epsilon}) \right]^{-1},
\end{align}
in agreement with Ref.~\cite{Braaten2006}. The diatom propagator is equivalent to the off-shell two-body T-matrix calculated in Eq.~\eqref{T2body} via:
\begin{equation}
\label{eq:Tdiatom}
    T(Z) = 8 \mu (2\pi)^3 g_2^2 D(Z).
\end{equation}
Thus by using the diatom trick, one can solve for and renormalize the two-particle scattering T-matrix exactly. \\

The Skorniakov–Ter-Martirosian (STM) equation for the three-body connected scattering amplitude can then be obtained from the Feynman diagrams shown in \ref{Fig1}b). The details of the calculation are presented in Ref.~\cite{Braaten2006}. Here we only present the final integral equation for the s-wave component of the three-body scattering amplitude:
\begin{align}\label{As_EFT}
     \mathcal{A}_s^{EFT}&\left( E, p, k\right)=\frac{16\pi}{a} \biggl[\frac{1}{2p k} \ln\left(\frac{p^2+p k+k^{2}-E}{p^2-p k+k^{2}-E}\right) + \frac{H(\Lambda)}{\Lambda^2}  \biggr]\nonumber\\
    &+ \frac{4}{\pi}\int_0^{\Lambda} dq\,q^2 \left(\frac{1}{2p q} \ln\left(\frac{p^2+p q+q^2-E} {p^2-p q+q^2-E}\right) + \frac{H(\Lambda)}{\Lambda^2}\right)\frac{  \mathcal{A}_s^{EFT}\left(E,q,k\right)}{-1/a+\sqrt{-E+3q^2/4-i\varepsilon}} .
\end{align}

We see that the main difference between Eqs.~\eqref{As_EFT} and \eqref{AC} is the determination of $H(E,p,q)$. In the effective field theory language, the role of $H(E,p,q)$ is played by the explicit three-body interaction $g_3 =-9 g_2^2 H(\Lambda)/\Lambda^2$, and is only a function of the UV cutoff $\Lambda$. While using separable potentials, we see there is no explicit need for an effective three-body interaction term, rather, $H(E,p,q)$ is completely determined by the two-body scattering properties. This comparison leads us to the effective form of the three-body interaction for the separable potential:
\begin{align}
   g_3 \underset{\Lambda\rightarrow\infty}{=} \frac{9g_2^2}{2}\int_0^\Lambda d\hat{\bf k}\frac{\xi\left(\left|{\bf k}+\frac{1}{2}{\bf p}\right|\right)  \xi\left(\left|{\bf p}+\frac{1}{2}{\bf k} \right|\right)-1}{p^2+{\bf p}\cdot{\bf k}+k^{2}-E}.
\end{align}

Before renormalizing the three-body scattering amplitude, first it is necessary to understand the behaviour of Eq.~\eqref{As_EFT} at large momenta, near the ultraviolet cutoff $\Lambda$. For simplicity, consider resonant two-body scattering, $a^{-1} = 0$, and large momenta, $p\to \infty$, while the remaining scales are in the infrared, $k,E\to 0$. If one neglects the inhomogeneous term proportional to $H(\Lambda)/\Lambda^2$, scale invariance implies the solution to $\mathcal{A}_s^{EFT}$ is a power-law in the momentum $p$ of the form \cite{bedaque1999,Bedaque1999renormalization,Braaten2006}:
\begin{align}
     \mathcal{A}_s^{EFT}\left( E, p, k\right)&\propto \frac{1}{p}\left[ A p^{i s_0}+ B p^{-i s_0}\right] \propto \frac{1}{p} \cos\left(s_0 \ln(p/\Lambda^*) + \phi_0\right),
     \label{scaling_solution}
\end{align}
where $s_0 = 1.00624$ and where $\Lambda^*$ is a constant with dimensions of momentum. The phase factor $\phi_0$ is irrelevant for our discussions and will be set to zero for the sake of simplicity.

The inhomogeneous term in the STM equation is needed to remove the UV divergences present in the solution to the three-body scattering amplitude. To explicitly fix the form of $H(\Lambda)$, 
we require that the STM equation is invariant under a change in $\Lambda$. To that end, let us consider an additional UV cutoff $\Lambda'>\Lambda$. 
We choose $H(\Lambda)$ in such a way that the three-body scattering amplitude at small momenta, $p,k \ll \Lambda$, is insensitive to the change in the UV cutoff. A direct comparison of the STM equation for the new cutoff $\Lambda'$ leads to:
\begin{align}\label{comparison_EFT}
    \left.\mathcal{A}_s^{EFT}\right|_{\Lambda'}& =\left.\mathcal{A}_s^{EFT}\right|_{\Lambda} +  \frac{4}{\pi}\int_{\Lambda}^{\Lambda'} dq \,q \left(\frac{1}{q^2} + \frac{H(\Lambda')}{\Lambda'^2}\right) \left.\mathcal{A}_s^{EFT}\right|_{\Lambda'} + \frac{4}{\pi}\left( \frac{H(\Lambda')}{\Lambda'^2} - \frac{H(\Lambda)}{\Lambda^2}\right) \int^{\Lambda} dq \,q \left.\mathcal{A}_s^{EFT}\right|_{\Lambda'},
\end{align}
where we have disregarded higher order corrections in $1/\Lambda$ that vanish when we take the cutoff to infinity.

The first term in Eq.~\eqref{comparison_EFT} is just the solution to the STM equation with a cutoff $\Lambda$, which we assume leads to the correct physics at low-energies. Thus we require the additional terms to vanish.
Substituting Eq.~\eqref{scaling_solution} into the additional terms and evaluating the integrals leads to the following condition on $H(\Lambda)$:
\begin{align}
    0 &= -\frac{\cos(s_0 \ln\left(\Lambda'/\Lambda^*\right) + \arctan(s_0))}{\Lambda'} + H(\Lambda')\frac{\cos(s_0 \ln\left(\Lambda'/\Lambda^*\right) - \arctan(s_0))}{\Lambda'} 
    - \left\lbrace \Lambda' \longleftrightarrow \Lambda\right\rbrace.
\end{align}
The solution can be readily obtained and yields the standard result \cite{bedaque1999,Bedaque1999renormalization,Braaten2006}:
\begin{equation}
    H(\Lambda) = \frac{\cos\left[s_0 \ln\left(\Lambda/\Lambda_*\right) + \arctan(s_0)\right]}{\cos\left[s_0 \ln\left(\Lambda/\Lambda_*\right) -\arctan(s_0)\right]}.
    \label{eq:H_EFT}
\end{equation}
Given $H(\Lambda)$, the low-energy properties of the three-body scattering amplitude will be independent of the choice of the cutoff $\Lambda$. 
However, it is still necessary to fix the scale $\Lambda^*$. In principle, this scale can be determined by matching the predictions of the EFT to more accurate calculations or experimental data. In this work, we match the scale $\Lambda_*$ to the wavenumber of the most deeply bound trimer state: 
\begin{align}\label{lambda*2}
    \Lambda_* = c \times\kappa_*^{(0)}\, e^{n\pi/s_0},
\end{align}
where $c$ is just a numerical constant. As also discussed in Sec.~\ref{section: comparative study}, we compare the predictions of the EFT to our separable potential method for the same value of $\Lambda$. We find that $c \approx 1.31$, while Ref.~\cite{Braaten2006} gives $c \approx 2.62$. 

\section{The two-channel model}
\label{sec:The two-channel model}
In the EFT treatment as presented in Sec.~\ref{subsec: EFT treatment}, we will show how the three-body scattering amplitude is most readily computed by performing a Hubbard Stratonovich transformation. Here, a diatom field is introduced and six-point terms in the Lagrangian are reduced to (maximum) four-point terms. The analogy between this approach and the treatment of the transition matrix using the separable expansion is most evident using a two-channel Feshbach model ~\cite{Taylor1972, Moerdijk1995, Timmermans1999,  Mies2000, Kokkelmans2002, Marcelis2004, Kholer2006, vandekraats2023}. In this model, we consider two free atoms that scatter in an (energetically) open channel $P$ and couple to a bound state in an (energetically) closed channel $Q$. This implies that the open-channel scattered state $\ket{\psi_P}$ satisfies the following Schr\"odinger equation 
\begin{align}
\label{eq:OpenCSchrodinger1}
E \ket{\psi_P} = (H^0_{PP}+V_{\mathrm{eff}})\ket{\psi_P},
\end{align}
where the effective potential operator $V_{\mathrm{eff}}$ corresponds to 
\begin{align}
V_{\mathrm{eff}} = V_{PP}+V_{PQ}G_{QQ}V_{QP},
\end{align}
with coupling strengths $V_{PQ} = V_{QP} = \alpha \ket{\xi}\bra{\xi}$ and closed-channel Green's function $G_{QQ}$. For a direct comparison with the EFT treatment presented in Sec.~\ref{subsec: EFT treatment}, we now proceed to set $V_{PP} = 0$, such that open-channel atoms only interact through direct coupling to the closed-channel. 
In the Feshbach formalism, this closed-channel is assumed to be energetically far removed from the zero-energy threshold and to contain only a single near-threshold bound-state. Under these assumptions, it is then possible to express the closed-channel Green's function $G_{QQ}(E)$ using the single-resonance approximation \cite{Marcelis2004}, such that 
\begin{equation} \label{eq:GQQ}
    G_{QQ}(E)=\frac{\ket{\phi_i}\bra{\phi_i}}{E-\nu},
\end{equation}
where $\nu$ is the energy of the bound-state (dimer) and where $\ket{\phi_i}$ represents a closed-channel eigenstate. Using the separable forms of $V_{PQ}=V_{QP}$ and implementing the single-resonance approximation, the Schr\"odinger equation for the open-channel scattered state $\ket{\psi_P}$ as introduced in Eq.~\eqref{eq:OpenCSchrodinger1} reduces to 
\begin{align}
\ket{\Psi_P} = \ket{\mathbf{k}}+  G^0_{PP} \ket{\xi}\frac{\alpha^2 \braket{\xi|\phi_Q} \braket{\phi_Q|\xi}}{E-\nu} \braket{\xi|\Psi_P},
\end{align}
where $G^0_{PP} = (E-H^0_{PP})^{-1}$ represents the free open-channel Green's function. Rewriting the previous expression to find a solution for $\ket{\Psi_P}$, and using this solution to compute the $T$-matrix from its definition $T\ket{k} = V_{\mathrm{eff}}\ket{\psi_P}$, we find that the two-channel transition matrix satisfies 
\begin{align}
\label{eq:Ttwochannel}
    \left\langle{\bf k'}\middle|T(E)\middle|{\bf k}\right\rangle& = \frac{\beta^2}{E-\nu- 4\pi\beta^2 \int_0^\Lambda dk \frac{k^2}{E-k^2/2\mu+i\epsilon} },
\end{align}
where we have introduced the rescaled coupling strength $\beta = \alpha \braket{\phi_i|\xi}$. 

The analysis of the transition matrix presented here is equivalent to the calculation of the transition matrix $T(E,\mathbf{q}) = g_2^2 D(E,\mathbf{q})$ as presented in Eq.~\eqref{eq:Tdiatom}, where $D(E,\mathbf{q})$ represents the dressed diatom propagator. The only difference is that, in the Feshbach formalism, the bare propagator in the closed-channel $G_{QQ}(E)$ equals $G_{QQ}(E) = (E-\nu)^{-1}$, whereas Eq.~\eqref{eq:lagrangian_Diatom} reveals that the bare propagator of the diatom $D_0(E,\mathbf{q})$ instead corresponds to $D_0(E,\mathbf{q}) = 4/g_2$.
Analogous to the treatment followed in Sec.~\ref{section: general framework}, the zero-energy limit of the transition matrix presented in Eq.~\eqref{eq:Ttwochannel} allows us to relate the rescaled interaction strength in the two-channel model $\beta$ to the two-body interaction strength $g_2$ in the EFT as 
\begin{align}
g_2 = \frac{- 4\mu \beta^2 (2\pi)^3}{\nu}.
\end{align}
Making this substitution in $T(E,\mathbf{q})/g_2^2$ together with the replacement $G_{QQ}(E) \rightarrow D_0(E)$ allows us to retrieve the diatom propagator as presented in Eq.~\eqref{eq:DiatomPropagator}.

\end{widetext}
\end{document}